# Optimizing Supply Chain Management using Gravitational Search Algorithm and Multi Agent System

**Muneendra Ojha**

Department of Information Technology, Indian Institute of Information Technology Deoghat, Jhalwa, Allahabad 211012, UP, India
muneendra@iiita.ac.in

**Abstract** Supply chain management is a very dynamic operation research problem where one has to quickly adapt according to the changes perceived in environment in order to maximize the benefit or minimize the loss. Therefore we require a system which changes as per the changing requirements. Multi agent system technology in recent times has emerged as a possible way of efficient solution implementation for many such complex problems. Our research here focuses on building a Multi Agent System (MAS), which implements a modified version of Gravitational Search swarm intelligence Algorithm (GSA) to find out an optimal strategy in managing the demand supply chain. We target the grains distribution system among various centers of Food Corporation of India (FCI) as application domain. We assume centers with larger stocks as objects of greater mass and vice versa. Applying Newtonian law of gravity as suggested in GSA, larger objects attract objects of smaller mass towards itself, creating a virtual grain supply source. As heavier object sheds its mass by supplying some to the one in demand, it loses its gravitational pull and thus keeps the whole system of supply chain perfectly in balance. The multi agent system helps in continuous updation of the whole system with the help of autonomous agents which react to the change in environment and act accordingly. This model also reduces the communication bottleneck to greater extents.

## 1 Introduction

Supply chain management can be defined as "a goal-oriented network of processes and stockpoints used to deliver goods and services to customers" [1]. Science of supply chains deals with an array of suppliers, plants, warehouses, customers, transportation networks and information systems that make up actual supply chains. But to study it we must break the overall complexity to a level which is relevant to the task at hand. We must define the application domain and then address the issues which are most important. In the above definition processes represent activities involved in producing and distributing goods or services. On the other hand stockpoints in this definition refers to the locations in supply chain where inventories are held. Finally processes and stockpoints are connected by a network describing various ways in which goods can be supplied through the chain. Supply chain has four levels of decision classification hierarchy - strategic, tactical, operational, and real time level, where strategic level is the most important among these, dealing with the resource assignment, goods movement, location of facilities to make it cost effective etc [2]. Success rate at other levels depend heavily on the decisions taken at this levels. However, the decision making is a difficult process as it involves the resolution of complex resource allocation and scheduling problem involving selection of different available stakeholders. According to the findings of Beamon [3], and Amiri [4], finding the best solution for supply chain management is NP-hard problem, so it must be strategically dealt with, on case to case basis developing efficient methodology that can find out the optimal or near-optimal solution in minimum computational time.

Over the last decades, research community has shown greater interest in analyzing natural phenomena and mapping their models to find out optimized solution for high-dimensional search space cases. Scientists have adopted various meta-heuristic methods like Genetic Algorithm [5], Simulated Annealing [6], Ant Colony Search Algorithm [7], Particle Swarm Optimization [8] etc. and shown that these algorithms are able to find efficient solution of complex computational problems like pattern recognition [9], image processing [10], filter modeling [11] etc. As like any other heuristic method, these algorithms do not promise to provide the most accurate and best solution for the problems but try to give the most optimized one, again on the case to case basis. So applying a new optimization algorithm on an untested domain is always an open problem.

In this paper we discuss the results of applying a relatively new swarm optimization algorithm, the Gravitational Search Algorithm (GSA) [12] in supply chain management domain. We have modeled the software application using multi agent system where intelligent agents autonomously take decision according to the changing environment and adapt to the perceived needs. In the following section we describe the relevance of GSA and Multi agent system in finding the solution of supply chain problem. Later we explain the problem structure with a mathematical model for the proposed clustering algorithm. Finally we end with implementation details and further scope of the research work.

## 2   Combining Multi Agent Technology with Swarm Optimization

Particle swarm optimization (PSO) is an evolutionary computational technique used for optimization motivated by the social behavior of individuals in large groups in nature [13]. Scientists and technicians have used different approaches to understand how such algorithms works and applied variations to optimize a large number of real life problems. They have been able to solve a variety of engineering problems, but fine tuning the PSO parameters as well as finding general criteria applied to all problems remains a major challenge. Thus we focus on a particular problem and try to find the best algorithm which is well suited to its requirements.

### *2.1   Why GSA is relevant in supply chain management problem?*

Swarm based algorithms use a collection of agents like ants or honey bees and perform parallel search with multiple starting points.  Each member carries out its intended operation with a target to maximize respective utility function, and shares its information with others. These operations are normally very simple, but their collective result produces astonishing results. This phenomenon in which local interaction between agents produce a global result facilitating the system to solve a problem without involving any central authority is known as swarm intelligence [14]. Gravitational Search Algorithm (GSA) is also a particle swarm optimization algorithm which works on the Newtonian laws of Gravity: "Every particle in the universe attracts every other particle with a force that is directly proportional to the product of their masses and inversely proportional to the square of the distance between them" [12]. It has found a good number of interested audiences albeit being relatively new. Just like other swarm intelligence techniques, it also models the laws of nature to find out an optimized solution to a non linear and dynamic problem domain. In the GS algorithm, each entity is considered an object whose performance is measured by the mass it carries. All objects attract each other by gravitational force which causes global movement towards objects of heavier mass. As heavier masses move slower than lighter ones, this creates virtual clusters representing better solutions. For the case of supply chain management, we assume all stakeholders including producer, consumer, supplier, warehouses, strategic partners, vendors, and haulers etc, as objects with some respective masses. We divide these stakeholder in two groups i.e. suppliers and customers. One can be supplier at some point of time and customer at other. Although we do not consider the demand part of the SCM process, if we focus on calculating the total cost of moving goods from a supplier to a customer, we can simulate a major part of the whole process of supply chain management. Cost of moving might include purchase price, cost of quality, inventory losses, transportation costs and any other factor which directly or indirectly affects the amount of money the customer has to shed. Thus suppliers act as objects of heavier masses or stars and customers imitate objects with lower masses, creating planetary systems revolving around stars.

## 2.2  Multi Agent System representing the complete supply chain

A rational agent is defined as an agent, which has clear preferences, models uncertainty via expected values, and always chooses to perform the action that results in the optimal outcome for itself from among all feasible actions [15]. A coordinated group of rational agents form the basis of multi agent system. Multi agent system is based on the assumption that agents coordinate among each other with the vision of achieving a common global goal. Referring the goal to be global is important here as individual agents have only local knowledge and thus local goals. But the agent system as a whole strives for one particular target to achieve. So whatever be the individual motives of agents, it should add an extra step towards the combined target of system and this is the raison d'être of these agents [16]. When an event happens each agent cooperates to organize a collaborative group to solve the problem. Agents communicate among the nearest of neighbors thus eliminating number of communications and form virtual clusters. In supply chain management (SCM) also, more and more consumers collect towards a prominent supplier thus forming some kind of demand supply cluster, where demands of consumers are fulfilled by nearest of supplier. Thus MAS presents an intuitive method of demand supply chain implementation in which each agent represents one or other real life entity participating in the whole process.

Many preceding studies have considered multi-agent system as the most suited way to solve complicated problems under the diverse environmental dynamisms [17][18]. A number of researchers have also successfully applied the multi agent technology for manufacturing, supply chain management, planning, scheduling, and control [19]. Their studies mainly focused on making better manufacturing system using an agent technology for automation and efficiency in the process planning and scheduling. Also, there are representative studies that have used MAS in developing intelligent manufacturing supply chain management systems e.g. the AARIA (Autonomous Agents for Rock Island Arsenal) project of Intelligent Automation company [20], ABCDE (Agent-Based Concurrent Design Environment) system developed by KSI (Knowledge Science Institute) of University of Calgary [21], a virtual manufacturing agent made by LIPS institute of the University of Texas at Austin [22] and MASCOT (Multi-Agent Supply Chain cOordination Tool) agent of Intelligent Coordination and Logistics team of Carnegie-Mellon University [23].

In the SCM problem, GSA with multi agents system is intuitively the most suited method of implementation. Every participating entity either a consumer or supplier is represented by an agent. Every agent carries some mass. All heavier agents are supplier and lighter ones are consumers. As the problem under consid-eration is the distribution among centers of Food Corporation of India (FCI), mass determines the total amount of food grains one center has in its stock. A center can be at the same time supplier for one type of agricultural produce (e.g. wheat) and consumer for some other type of grains (e.g. rice). Thus effective mass and gravitational pull is context dependent.

## 3   FCI distribution system and methodology

Food Corporation of India (FCI) was established to fulfill the objectives of procurement of agricultural products, its storage, movement, export if asked, disposal in case of excess amount, and quality control. FCI has five zones with each zone having its member state representing region.

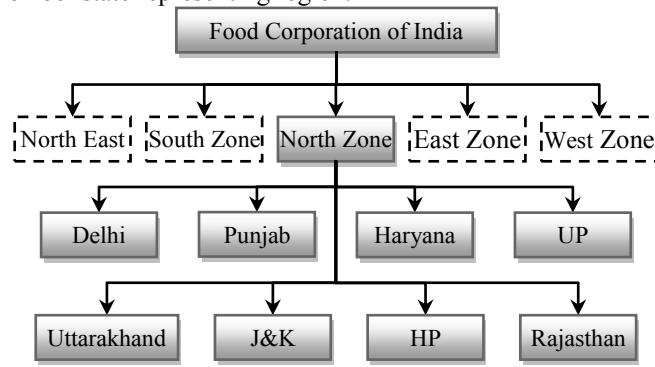

**Fig. 1.** Hierarchical structure of FCI collection and distribution centers. North zone itself has 54 district centers in total involved in the procurement and other food management activities.

For example, North zone of FCI has eight regions namely Delhi, Haryana, HP, J&K, Punjab, Rajasthan, UP and Uttarakhand regions as depicted in the Figure 1, which in turn have 54 district centers. Punjab & Haryana Regions are the major surplus states as far as production/procurement of food grains is concerned. More than 90% stocks of these supplier states are moved to other parts of the country [24]. Apart from FCI the other major agencies involved in storage and warehousing of food grains in public sector are Central Warehousing Corporation (CWC) and 17 State Warehousing Corporations (SWCs). For procuring food grains from farmers, FCI creates nodal centers where it purchases at a price fixed by the Government. It is then sent to respective region's storage centers or warehouses. Every warehouse has a minimum requirement as well as maximum capacity.

With the broad requirements clear about FCI and its supporting agencies, we can model the whole distribution system as MAS. An agent in our system represents a district center of FCI. As there are 54 district centers in North Zone, so there will be 54 agents corresponding to them. Each agent thus would have a minimum requirement, i.e. a minimum amount of particular food grain, which it has to maintain as reserve at any given instance. This reserve is decided on many factors like, Government of India policy on food security, type of food, population of the region, requirements for the region in near future, rate at which food is being consumed in public distribution system, state rules etc. Any amount of food above this reserve level is assumed to be surplus which can be sent to other agents in the system if required. Thus we can clearly differentiate between supplier and consumer at any point of time using following definition:

**Supplier**: *An agent which has food grains in excess of its set reserve amount and any point of time.*

**Consumer**: *An agent which has food grains below the decided reserve at an instance of time.*

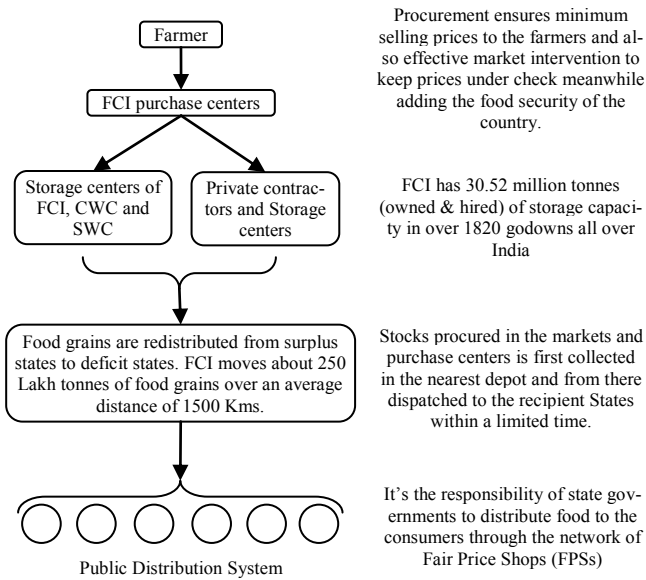

**Fig. 2.** Broad functionality and objectives of FCI and its supporting agencies like Central Warehousing Center, State Warehousing Center and Public Distribution Centers. Source: Food Corporation of India [24]

As this definition is dependent on time and type of food grain, any agent can be a supplier for one type and consumer for other type of food grains at the same point of time. Two reasons can be visualized for automatic redistribution of food grains between centers: first, when a center needs it because it has receded below its reserve or when a center has accrued more than what it can efficiently manage. Both of these cases prompt their representative agents to autonomously act in case such scenario is perceived in the environment. Though FCI has its structural limitations and boundaries, we assume it to be free from any such regulations. Although we do suggest a priority queue for suppliers in which higher priority must be given to consumers of the same zone or state rather than a consumer of differ-ent zone. Figure 2 gives a broad idea of the levels at which different agencies work and how food grains purchased from farmer reaches household.

## 4 The mathematical model

Each participating physical entity in this supply chain system is represented by an agent. There are agents for zonal offices, regional states, district centers, nodal centers, storage warehouses, and purchase centers. Out of these, any agent can be supplier or consumer depending on the resources it owns. A consumer is consumer till it has requires food grains to achieve a particular target level referred to as set reserved limit. It is decided by an array of parameters namely, type of food grain, near future requirement prediction, number of Public Distribution Centers (PDCs) it caters, geographical area this agent covers, positional importance, and minimum food reserves it has to maintain (decided as a policy matter by government, hardcoded but can be changed if needed).

In GSA, initially the system is considered a randomly distributed isolated space of masses (agents). Each agent carries some mass and in gravitational system, every object gets attracted by gravitational pull of another object. In our system we do not assume an agent to move. Rather we assume that whichever supplier exerts greatest force of attraction on consumer will be the first one to attempt in fulfilling its requirements.

At a time 't' we define force acting on a consumer mass '$M_c$' from a supplier mass '$M_s$' to be

$$F_{cs}(t) = -G_c(t)\frac{M_c(t) \times M_s(t)}{R} \qquad (1)$$

Where,

- $F_{ij}(t)$ = Force of attraction by a supplier and consumer at time t.

- $M_c(t)$ = Mass of Consumer at time t
  = Amount it can supply
  = Reserve limit – Current Stock

- $M_s(t)$ = Mass of Supplier at time t
  = Amount it needs
  = Reserve limit – Current Stock

- $G_c(t)$ = Urgency factor, initially set randomly between 0 to 1. Increases with time.

- $R$ = Possible cost incurred in supplying food grains per unit to the consumer.

An agent can be attracted by another agent by a positive force of attraction. Putting a minus sign in equation 1 ensures that no two consumers or suppliers are attracted towards each other.

Unlike the traditional gravitational search algorithm proposed by [12] we have made some changes in our algorithm. We consider the attracting force to be one dimensional only. We were thus able to eliminate the possibility in which multiple smaller suppliers could collaborate and exert greater force than a single large supplier. In this case combined force could take a consumer to cluster of smaller suppliers but do not consider the cost incurred in multiple smaller transportations. We have also set R to be the cost incurred per unit where,

$$F \propto 1/R \quad (2)$$

As cost increases attraction between supplier and consumer decreases though it never becomes zero as is the case in gravitational forces of nature as well. Gravitational constant $G_c$ is initially randomly set as the urgency factor defining how urgent is the need of this particular customer. As food is supplied to it and need decreases, value of $G_c$ also changes with time. Thus it is a function of initial value $G_0$ and time t:

$$G(t) = f(G_0, t) \quad (3)$$

Masses of individual agents are also dependent on time t. As supply starts, the total available stock of both consumer and supplier change. This changes the respective masses and thus mutual force of attraction is also decreased. If there are N agents, position of an agent is defined in 2-dimensional space as:

$$X_i = (x_i^1, x_i^2) \; for \; i = 1, 2, \ldots, N \quad (4)$$

When agents attract each other they tend to move towards each other in the virtual space. This movement is not reflected in the real scenario nor has any physical significance. Maximum closeness between two agents during clustering phase is restricted to sum of their radii.

## 5  Implementation and result discussion

Application simulation is divided in two phases. First one is the clustering phase when we form virtual clusters of supplier consumer agents depicting some kind of planetary system in which supplier is the star and consumer are the planets revolving around it. As the clustering phase starts, agents get attracted towards those of greater mass and their respective Euclidean distance decreases. When clustering ends each supplier is attached with many consumers as shown in Figure 3. If the supplier has enough capability to fulfill demands of more than one con-sumer it must decide which consumer to cater first. Thus every supplier creates a priority queue on the basis of first hierarchical boundaries of FCI and then in the decreasing order of consumers' urgency factor. Hierarchical boundaries are important here as we assume an agent will cater to the needs of its state prior to sending grains to neighboring state. Providing autonomy to the agent system, we assume that if a consumer finds a supplier 'A' in neighboring state providing cheaper food than agent 'B' in the same state it will approach 'A' rather than 'B' as long as both states are in same zone of FCI.

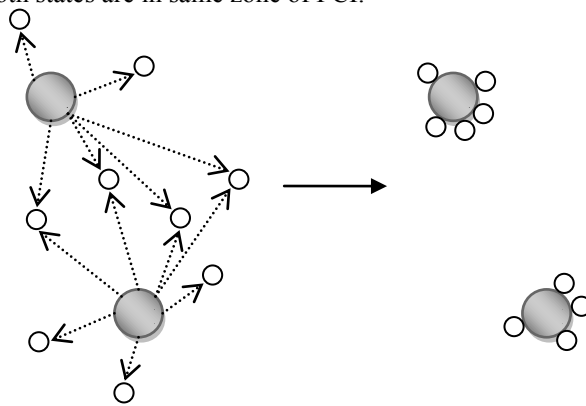

**Fig. 3.** Initially all agents are scattered in the search space and are attracted by more than one supplier. But they move closer to the one which has greater force of attraction forming virtual clusters.

Actual supply of goods can only be realized physically and once it is approved and done, agents' information must be updated manually. As the system itself notifies, how much food is to be provided by a supplier to the consumer; it can automatically update its inventory after approval. Once the inventory is updated, effective mass of both supplier and consumer changes. Agents that were customer for some time but their needs have been satiated will move out of the SCM process and similarly an agent that was a supplier for some time but its excess mass is already shed, will also exit the system. This behavior is perfectly represented by multi agent system as MAS gives autonomy and intelligence to its agents to participate in the process as long as required or leave the environment whenever deemed necessary.

There is no need of a centralized command to dictate the centers which must be catered. Agents automatically form collaborative supplier-consumer clusters and cater to the needs of each other thus removing the bottleneck of central node. Number of consumers and suppliers as a given time can be generated using Poisson distribution. As soon as a center becomes consumer the representative agent passes on its identification and requirement message to its immediate neighbors which in turn pass it on to their respective neighbors and so on. With the one hop communication channel, agents do not need to broadcast requirement to whole network thus reducing burden on the network. All the supplier agents respond back with their identification and availability. When consumer receives response of a supplier it refers to the database in order to fetch supplier information like location, state, zone, modes of transportation, price of grain and any other constraints based on which the final cost is calculated. It calculates the attraction force using equation 1 and updates this agent as the global best. A consumer also maintains list of suppliers for as soon as new supplier comes up, the queue is reshuffled and new global best is updated if required. In case one supplier is not able to fulfill the requirement, consumer will approach the next in queue without pronouncing its need again on the network.

## 6 Conclusion

The current system can be implemented for very large scale real life demand supply chain management. This system can be applied to observe the pattern of product usage, to find out high demand zones, or the consumption rate of an area. It can also draw out the reliability factor of a supplier and cost to benefit ratio for a product. This system can also help organizations in evaluating needs of customer and adapt accordingly.